\documentstyle[epsfig,a4,12pt]{article}
\epsfverbosetrue

\def\nuebar{\bar{\nu_e}}
\def\dm2{\rm{\Delta m^2}}
\def\munu{\mu_{\nu}}
\def\nurad{\rm{ < r^2 > }}
\def\s2tw{\rm{ sin ^2 \theta _W }}
\def\am241{\rm{ ^{241} Am }}
\def\u238{\rm{ ^{238} U }}
\def\th232{\rm{ ^{232} Th }}
\def\k40{\rm{ ^{40} K }}

\begin{document}

\hfill AS-TEXONO/98-04

\begin{center}
\large
\bf{
A Pilot Experiment with Reactor Neutrinos 
in Taiwan \\
}
\vspace*{0.5cm}
\normalsize
Henry T. Wong~$^{\alpha}$  and Jin Li~$^{\beta}$  \\
\vspace*{0.5cm}
$^{\alpha}$ Institute of Physics, Academia Sinica, Taipei, Taiwan. \\
$^{\beta}$ Institute of High Energy Physics, Beijing.
\end{center}
\vspace*{0.5cm}
\begin{center}
{\bf
Abstract
}
\end{center}

A Collaboration comprising Taiwan and mainland Chinese scientists has been
built up since 1996 to pursue a experimental program in neutrino and
astro-particle physics in Taiwan.
A pilot experiment to be performed at the Nuclear Power Station II
in Taiwan is now under intense preparation. 
It will make use of 
a 600~kg CsI(Tl) crystal calorimeter to 
study various neutrino interactions.
The feasibility of performing a long baseline reactor neutrino
experiment will also be investigated. 
The conceptual design and the physics to be addressed 
by the pilot experiment are presented.

\section{Introduction}

Activities in experimental particle physics started in
Taiwan in the late 80's. 
There is intensive participation in a number
of international experiments (L3, CDF, Fermilab E871, Belle,
AMS, RHIC-PHOBOS), working on various hardware, online
software and data analysis projects.
A neutrino group was started in 1996, looking
into the possibilities and opportunities in
reactor neutrino experiments to be performed
in Taiwan~\cite{win97}.

At present, the ``TEXONO''~\footnote{
{\bf T}aiwan {\bf EX}periment {\bf O}n Reactor {\bf N}eutrin{\bf O}.}
Collaboration comprises more than 40 scientists
with diversified expertise
from Taiwan (Academia Sinica, Institute of Nuclear
Energy Research, National Taiwan University, National Tsing Hua
University, 
National Chiayi Teachers' College
and Nuclear Power Plant II), 
mainland China (Institute of High Energy Physics,
China Institute of Atomic Energy, 
Nanjing University,
Shandong University, University of Science and Technology at Hefei)
and the United States (University of Maryland).
It is one of the first collaborative efforts in large-scale
basic research among Taiwanese and mainland Chinese scientists.

The principal objectives of the Collaboration are to 
initiate a program in experimental neutrino physics
and astro-particle physics, as well as to 
build up a qualified experimental team for future
projects.
The goal is to conduct 
an international-standard particle
physics experiment in Taiwan.
A good local research program is complementary
to the participation of international projects,
and is essential to the solid foundation of
international collaborative basic research in the
country. 
A local program can serve as 
a base for training young students and scientists,
and a ``launch-pad'' for more ambitious project
beyond, local or abroad.

\begin{figure}
\centerline{
\epsfig{file=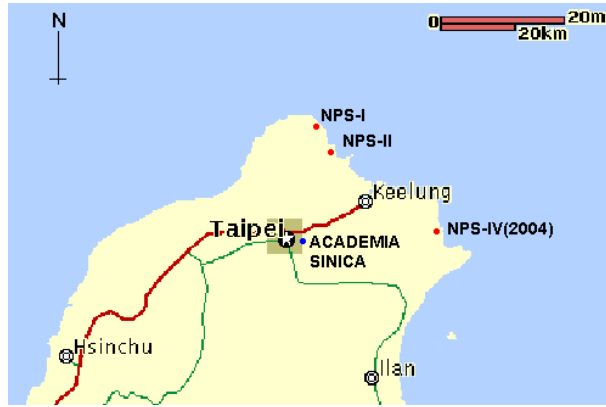,width=8cm}
}
\caption{
Sketch map of the northern shore of Taiwan, showing the
three nuclear power stations and their relative position
with Taipei city. The Pilot Experiment will be performed
at NPS-II as indicated.
}
\label{taipeimap}
\end{figure}

Given the above implications and considerations,
the field of choice is reactor neutrino.
There are operational power reactors in
Taiwan with parameters shown in Table~\ref{reactor}.
The mountainous landscape  dotted with mines and tunnels
makes the construction of an underground laboratory
conceivable. The proximity of the reactor locations 
and the possible underground sites (Nuclear Power 
Plants I, II and IV are all about 20-30~km from 
Taipei city, as shown in Figure~\ref{taipeimap}) 
to the city infrastructures provides
an additional advantage. 

\begin{table}[t]
\caption{Power Reactor Plants in Taiwan and their Parameters
\label{reactor} }
\begin{center}
\begin{tabular}{|l|l|c|c|l|l|} 
\hline
Plant & Type & No.& Power & Location & Status \\  
& & Cores & /core~GW & & \\ \hline \hline
I & Boiling Water & 2 & 1.78 & North Shore & Operational \\
II & Boiling Water & 2 & 2.90 & North Shore & Operational \\ 
III & Pressurized Water & 2 & 2.78 & South Shore & Operational \\
IV & Advanced Boiling Water & 2 & 4.1 & North Shore 
& Expect 2004 \\ \hline
\end{tabular}
\end{center}
\end{table}

\begin{figure}
\centerline{
\epsfig{file=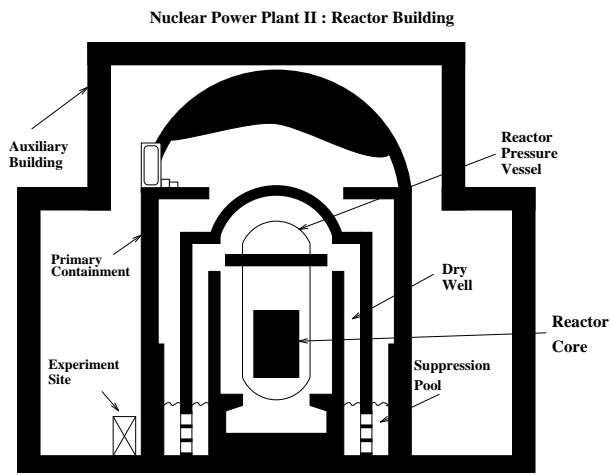,width=8cm,angle=270}
}
\caption{
Schematic side view, not drawn to scale,
of the NPS-II Reactor Building,
indicating the experimental site.
The reactor core-detector distance is about
30~m.
}
\label{fplanside}
\end{figure}

The Collaboration has been intensely 
preparing a ``pilot'' experiment to be 
performed at a site of about 30~m from
one of the reactor cores at Nuclear Power Station II.
The location is shown schematically in Figure~\ref{fplanside}.
Meanwhile, the feasibility and conceptual studies of the ``next''
project will be pursued.
A possible direction can be a long baseline reactor
neutrino experiment.

\section{The Pilot Experiment}

\subsection{Considerations and Constraints}

The most important subject
in the field of reactor
neutrino physics at present 
is undoubtedly a long baseline
neutrino oscillation experiment~\cite{rnuphys}.
The next generation
experiment will have to be at a
location of at least
O(10~km) from the reactor cores at
a depth of at least 500~m of rock underground
and with a detector target  mass of O(1000~ton).
Taiwan does have the geographical advantage
to perform such an experiment. However,
the magnitude of resources required (in
terms of funding, expertise, manpower, time
and international credibility) to launch
such an experiment is considered to be too
large as the first project of
a starting group. It would
be more appropriate to first pursue a smaller
experiment located close to the reactor
cores.

Such a  ``pilot'' experiment will be an
important learning exercise and team-building efforts
for the new Collaboration, preparing ourselves
for the more ambitious projects beyond,
whatever exactly they may be.
In addition, it should have its stand-alone and
independent physics motivations, while
being the first efforts towards a long-term
visions. 
It is essential that the experiment itself will
produce interesting and new physics
results, with a detector that can be built,
and hence the production of physics output possible,
in a relatively short time.

\subsection{Physics and Detector Motivations}

Almost all previous reactor neutrino experiments were
based on liquid scintillator techniques to
study the ($\rm{ \bar{\nu_e}~p  }$) ``Reines'' interactions,
but with different neutron-capture isotopes.
An experiment focusing on 
gamma detection has never been attempted.

However, gamma-ray spectroscopy has been a
standard technique in nuclear sciences
(that is, in the investigations of physics
at the MeV range). Gamma-lines of characteristic
energies give unambiguous information on the
presence and transitions of whichever isotopes,
allowing a unique interpretation of the physical
processes.
The experimental difficulties of building a
high-quality gamma detector for MeV neutrino
physics have been the large target mass required.
However, in the past few years, big electro-magnetic
calorimeter systems (with mass up to 40 tons of
crystals, in the case
for the forthcoming B-factories experiments~\cite{bfactories})
have been built for high energy physics experiments,
using CsI(Tl) crystals with photo-diodes readout.

\begin{figure}
\centerline{
\epsfig{file=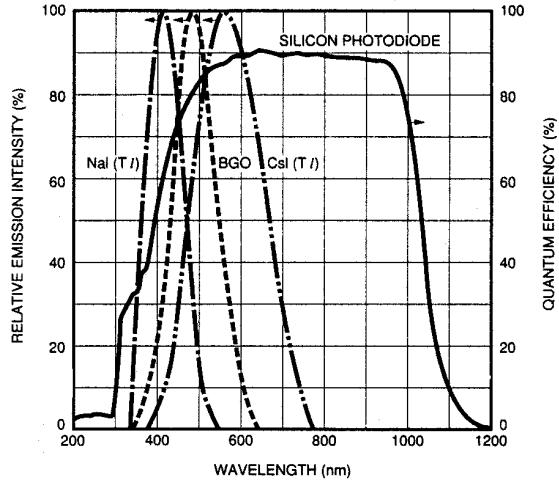,width=8cm}
}
\caption{
The sensitivities as a function of
wavelength for typical silicon
photo-diode, and the emission
spectra of several common crystal scintillators,
showing that CsI(Tl) matches best among them.
}
\label{pdspec}
\end{figure}

The properties of
CsI(Tl) crystals, together with those
of a few common scintillators, are listed in
Table~\ref{scintab}. The CsI(Tl) crystal offers
certain advantages over the other possibilities.
It has relatively high light yield
and high photon absorption
(or short radiation length). It is mechanically
stable and easy to machine, and is only
weakly hygroscopic. It emission spectra well matches
the response of silicon photo-diode
as depicted in Figure~\ref{pdspec}, thus making
a compact design with minimal passive volume
and efficient shielding configuration
possible.

\begin{table}
\begin{center}
\begin{tabular}{|l|c|c|c|c|c|c|}
\hline
Properties & CsI(Tl) & NaI(Tl) & BGO & Liquid & Plastic 
& Glass \\ \hline \hline
Density & 4.51 & 3.67 & 7.13 & 0.9 & 1.0 & $\sim$3.5 \\
Relative Light Yield & 0.45 & 1.00 & 0.15 & 0.4 & 0.35 
& 0.15  \\
Radiation Length (cm) & 1.85 & 2.59 & 1.12 & $\sim$45 & $\sim$43 & 4 \\
Emission Peak (nm) & 565 & 410 & 480 & 425 & 425 & 395 \\
Decay Time (ns) & 1000 & 230 & 300 & 2 & 2 & 100 \\
Refractive index & 1.80 & 1.85 & 2.15 & 1.5 & 1.6 & 1.55 \\
Hygroscopic & slightly & yes & no & no & no & no \\ \hline
\end{tabular}
\end{center}
\caption{Characteristic properties of the common
crystal scintillators and their comparison with
typical liquid, plastic and glass scintillators.}
\label{scintab}
\end{table}

The CsI-crystal production technology is by now well matured
and the cost has been reduced enormously due to the
large demands. It become realistic and affordable
to build a CsI detector in the range of 1-ton
in target mass
for a reactor neutrino experiment.
The detector mass 
can be further scaled up 
if the first experiment would
yield interesting results or lead to
other potential applications.

Since a crystal
calorimeter with improved $\gamma$-detection capabilities
is a new approach to neutrino experiment, 
it does offer 
several physics opportunities
with a rather modest investment of cost, manpower and time,
well suited for a starting group. It is also technically
much simpler to build and to operate than, for instance,
gas chambers and liquid scintillators.
The background processes for such novel applications
are not thoroughly understood. It is a new research domain the
experiment will explore and will require state-of-the-art
techniques in its control, identification and suppression.

\subsection{Physics Menu}

Previous experiments with reactor neutrinos
primarily focused on the
($\rm{ \bar{\nu_e}~p  }$) interactions
to look for neutrino oscillations~\cite{refrnuexpt}.
However, the use of low energy (MeV) neutrino
as a probe to study particle and nuclear physics
has not been well explored - although high
energy (GeV) neutrino beams from accelerators
have been very productive in investigating electroweak,
QCD and structure function
physics~\cite{nuint} and have blossomed into
a matured field. There are rooms for interesting
physics with reactor neutrino experiments along this
direction, some of which can be explored
by a crystal calorimeter.

\subsubsection{Neutrino-Electron Scattering}
\begin{figure}
\centerline{
\epsfig{file=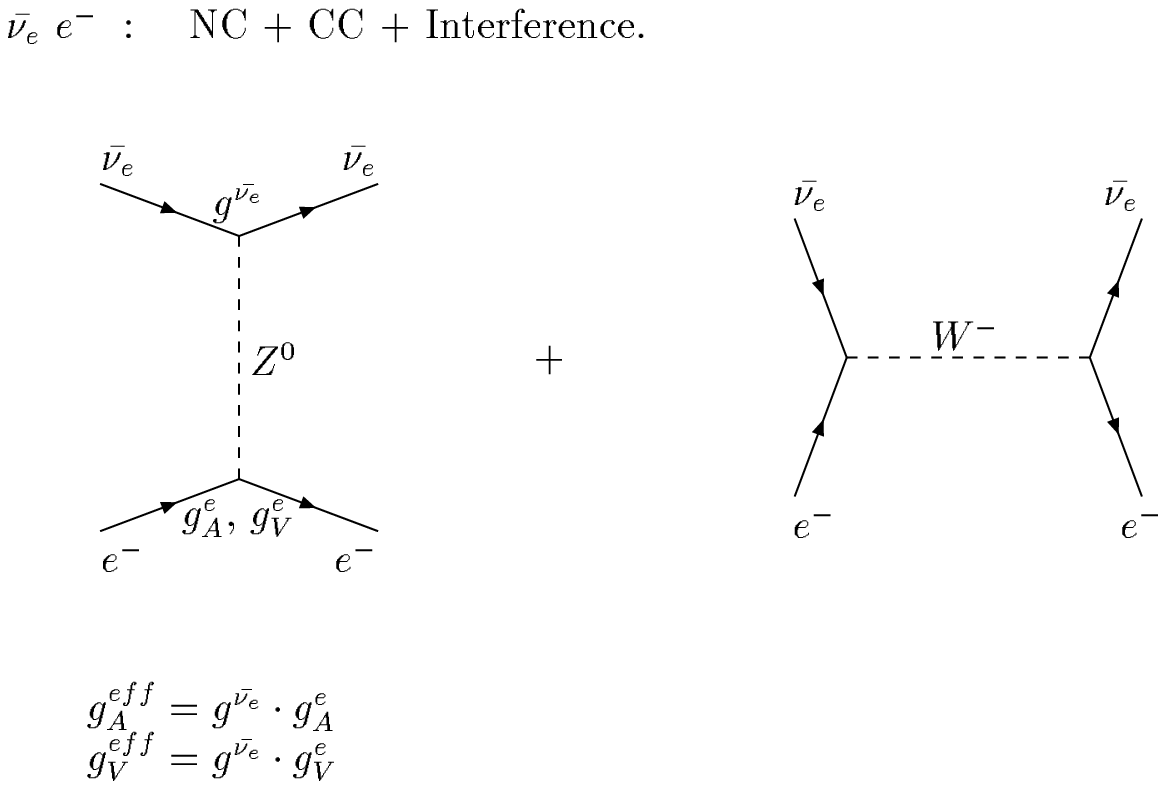,width=10cm}
}
\caption{
Feynman diagrams for ($\nuebar$ e) interaction,
showing there exists charged and neutral current
components, as well as their interference term.
}
\label{nuefeynman}
\end{figure}

The cross section
for the process
\begin{displaymath}
\rm{
\bar{\nu_e} ~ + ~ e^- ~ \rightarrow ~ \bar{\nu_e} ~ + ~ e^- 
}
\end{displaymath}
gives information on the electro-weak parameters 
($\rm{ g_V , ~ g_A , ~ and ~ sin ^2 \theta_W }$), 
and are sensitive to 
small neutrino magnetic moments ($\munu$) 
and the mean square charge 
radius ($\nurad$)~\cite{nueth}.
Scatterings of the 
$\rm{( \nu_e ~ e )}$ and
$\rm{( \nuebar ~ e )}$ are two of the most
realistic systems
where the interference
effects between Z and W exchanges 
can be studied,
as shown by the Feynman diagrams in Figure~\ref{nuefeynman}.

In an experiment, what can be measured is the recoil
energy of the electron (T). The differential cross
section can be expressed as :
\begin{eqnarray*}
\frac{ d \sigma }{ dT } ( \nu ~ e ) & =  &
\frac{ G_F^2 m_e }{ 2 \pi } 
[ ( g_V + x + g_A )^2 + ( g_V + x - g_A )^2  [ 1 - \frac{T}{E_{\nu}} ]^2
+ ( g_A^2 - ( g_V + x )^2  ) \frac{ m_e T }{E_{\nu}^2}  ] \\
 &  &  + \frac{ \pi \alpha _{em} ^2 \munu ^2 }{ m_e^2 }
[ \frac{ 1 - T/E_{\nu} }{T} ]
\end{eqnarray*}
where 
$ g_V = 2 ~ \s2tw - \frac{1}{2}$ and $ g_A =  - \frac{1}{2}$
for $ \nu_{\mu} ~ e$ and $\nu_{\tau}~ e$ scatterings
where only neutral currents are involved,
and
\[
x = \frac{2 M_W^2}{3} \nurad \s2tw ~~ \rm{for} ~~ \nu ,
\]
while replacing
\[
\left. \begin{array}{c}
g_A \rightarrow - g_A \\
x \rightarrow - x 
\end{array}
\right. ~~ \rm{for}  ~~ \bar{\nu} ~ .
\]
For $\nu_e ~ e$ scattering, both neutral and charged currents
and their interference terms contribute, so that the cross sections
can be evaluated by replacing 
$g_V \rightarrow g_V + 1$ and 
$g_A \rightarrow g_A + 1$.

\begin{figure}
\centerline{
\epsfig{file=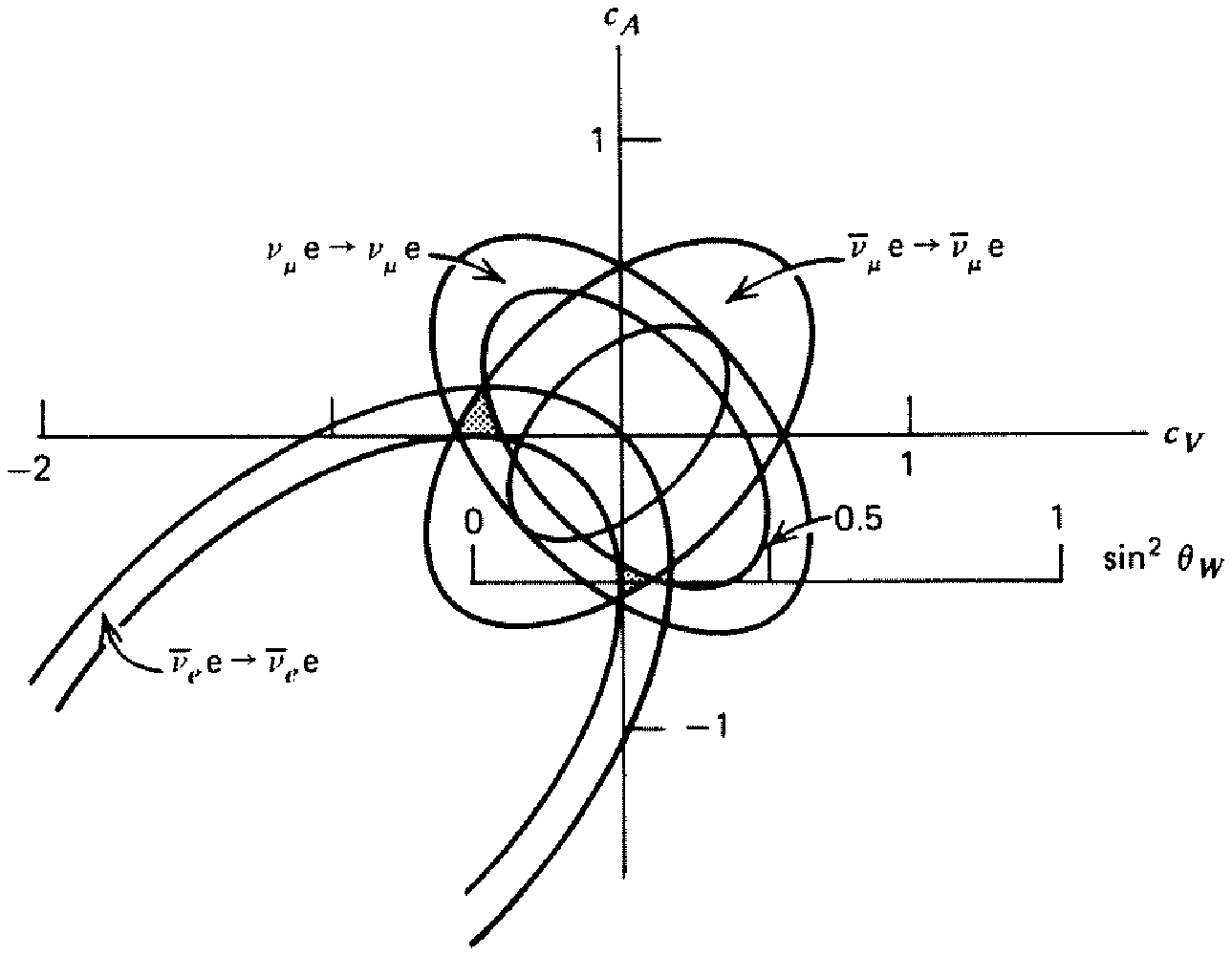,width=7cm}
}
\centerline{
\epsfig{file=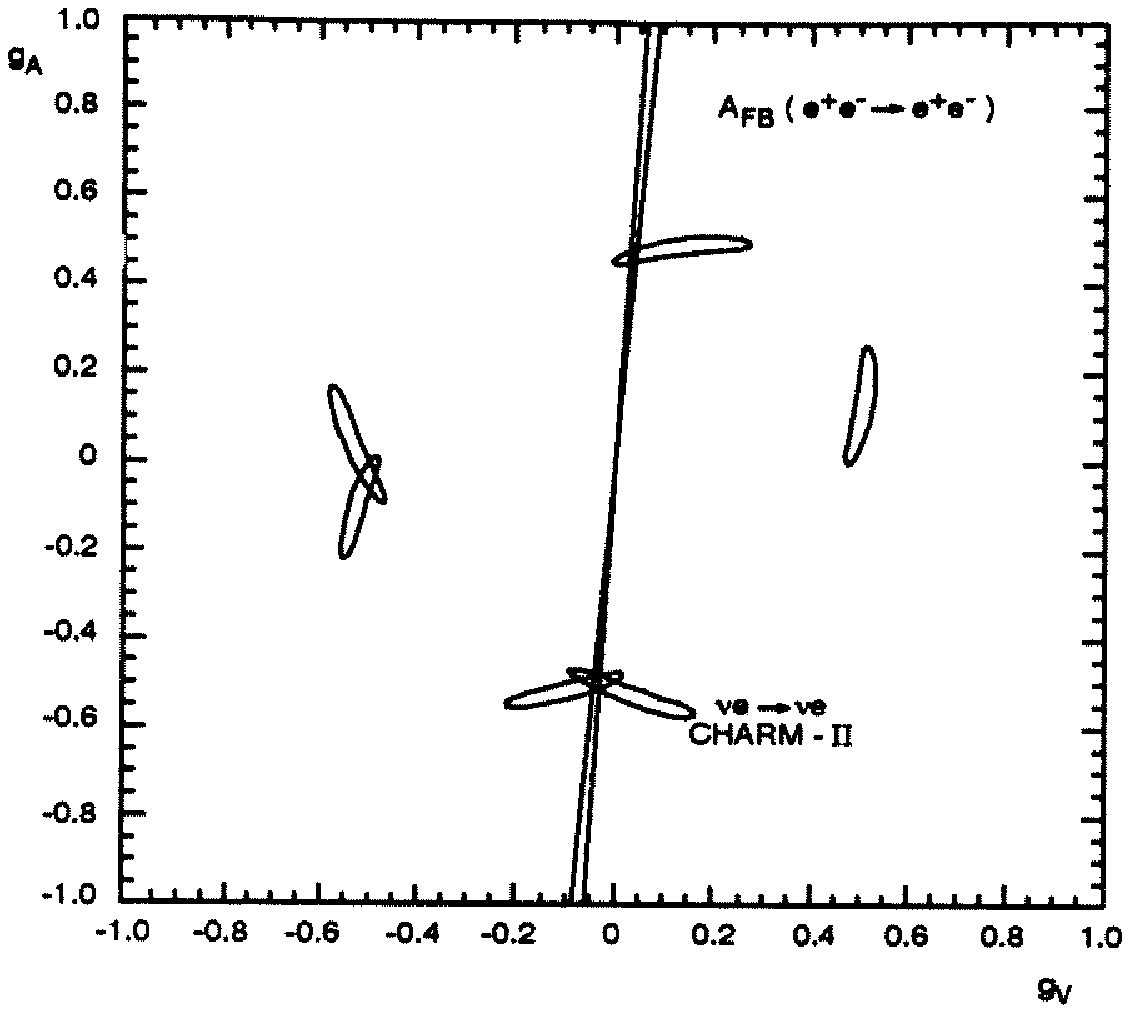,width=7cm}
}
\caption{
{\bf Top} : Sensitivities of ($\nu_{\mu}$ e),
($\bar{\nu_{\mu}}$ e) and ($\nuebar$ e) cross-sections to
different regions in the $\rm{g_A^e}$-$\rm{g_V^e}$ parameter
space (axes labeled  as $\rm{c_A}$ and $\rm{c_V}$), showing
their complementarity.
{\bf Bottom} : Allowed region of the
$\rm{g_A^e}$-$\rm{g_V^e}$ plane
from the accelerator ($\nu$ e) scattering  and
the $\rm{e^+ e^-}$ forward-backward symmetry data.
}
\label{gvvsga}
\end{figure}

\begin{figure}
\centerline{
\epsfig{file=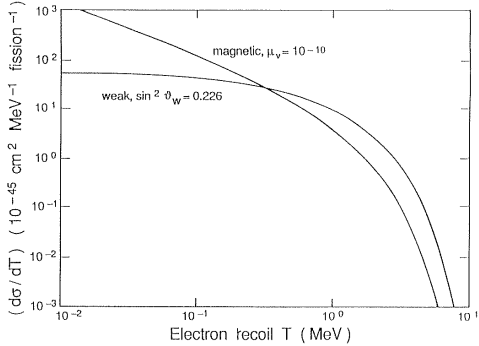,width=7cm}
}
\caption{
Differential cross section showing the
electron recoil energy spectrum in
$\nuebar$-e scatterings, for
of Standard Model processes and
for the case with a neutrino
magnetic moment of 10$^{-10}$ Bohr magneton,
the present experimental limit.
}
\label{nuerecoil}
\end{figure}

The $\rm{g_A^e ~ Vs. ~ g_V^e}$ parameter space where
$\rm{( \nuebar ~ e )}$ scatterings are sensitive to
is depicted in Figure~\ref{gvvsga}. The complementarity
with $\rm{ ( \nu_{\mu} ~ e , ~ \bar{\nu_{\mu}}  ~ e ) }$ 
can be readily seen. 
The expected recoil energy spectrum 
is displayed in Figure~\ref{nuerecoil}, showing
standard model expectations and the case with
an anomalous neutrino magnetic moment at the
present limit. The $\munu$ term have a 
$\rm{\frac{1}{T}}$ dependence. 
Accordingly, experimental searches for
the neutrino magnetic moment should 
focus on the reduction of the threshold
(usually background-limited) for 
the recoil electron energy.

Therefore, investigations of $\rm{( \nuebar ~ e )}$
cross-sections with reactor neutrinos
allow one to study electro-weak physics 
at the MeV range, to probe charged and neutral
currents interference,
and to look for an anomalous neutrino magnetic moment.

A 600~kg CsI crystal calorimeter will have more
target electrons than previous 
experiments~\cite{nueexpt} and 
current projects~\cite{nuenew},
and thus can potentially improve the
sensitivities of these studies.
The compact detector size will also
allow effective shielding design.
The signature for  $\rm{( \nuebar ~ e )}$
will be a single hit 
out of the several hundred channels
in the active target configuration.
The goal is to achieve a 10\% measurement
on the cross-section and a physics
threshold of 1~MeV to probe neutrino
magnetic moment down to 
$\rm{5 \times 10^{-11} ~ \mu_{B}}$

\subsubsection{Neutrino Charged and Neutral Currents on Deuteron} 
The interactions 
\begin{displaymath}
\rm{
CC ~ : ~ \bar{\nu_e} ~ + ~ d ~ \rightarrow ~
e^+ ~ + ~ n ~ + ~ n ~
}
\end{displaymath}
and
\begin{displaymath}
\rm{
NC ~ : ~ \bar{\nu_e} ~ + ~ d ~ \rightarrow ~
 \bar{\nu_e} ~ + ~ p ~ + ~ n 
}
\end{displaymath}
have been observed~\cite{nudexpt},
and calculated~\cite{nuex,nudth}.
Improved measurements will be of interest, especially 
since the NC reaction is the detection channel adopted
by the forthcoming SNO experiment~\cite{sno} 
for solar neutrino detection. 
Measurement of the CC/NC ratio provides a
complementary method 
to search for neutrino oscillations, which
is independent of the
detailed knowledge of the neutrino
source - an interesting possibility for long-baseline
experiments which may receive neutrinos from many
reactor cores and where the conventional
``Reactor ON-OFF'' subtraction may not be feasible.
The SNO experiment will pursue this
CC/NC ratio measurements for solar neutrino,
and it would be desirable to have a
laboratory experiment exploring 
the systematics of the technique.

In a realistic experiment, the 
CsI crystal slabs will be put into a tank with
500~kg of heavy water ($\rm{ D_2 O }$).
Neutrons produced will mostly be captured
via (n,$\gamma$) by $^{133}$Cs and $^{127}$I.
The CC signatures will be rather spectacular: 
two back-to-back
511~keV $\gamma$s followed by two separate bursts of 
high energy $\gamma$s produced in 
neutron capture.  The
NC detection will rely on a single $\gamma$-burst.

This is a complementary -  and improved -
technique to the previous
experiments~\cite{nudexpt}
which used $^3$He proportional counters
and were therefore sensitive only to neutrons.
Accordingly, the CsI detector,
with its $\gamma$-detection capabilities,
can differentiate the signals
from the other neutron-producing background channels :
\begin{eqnarray*}
\rm{ ``Reines" ~ : ~ \bar{\nu_e} ~ + ~ p ~ }  & \rightarrow ~ &
\rm{ e^+ ~ + ~ n ~ (Threshold = 1.80~MeV) ~~ and } \\
\rm{ \gamma ~ Dissociation ~ : ~ ~
\gamma ~ + ~ d ~}  & \rightarrow ~ &
\rm{ p ~ ~ + ~ n  ~  (Threshold = 2.23~MeV) } ~,
\end{eqnarray*}
and can prevent the CC events with one undetected
neutron from contaminating the NC sample.
The goals are to achieve  5\% and 10\%
measurements for $\nuebar$d-CC and $\nuebar$d-NC,
respectively.

\subsubsection{Neutral Current Excitation on $^{10}$B
and $^{11}$B }   
If a compact boron-rich object
like B$_4$C (natural boron consists
of 20\% $^{10}$B and 80\% $^{11}$B)
is used as the passive target, 
characteristic $\gamma$-lines 
(3.59, 5.16~MeV for $^{10}$B, and 
2.11, 4.45, 5.02~MeV
for $^{11}$B) 
will be emitted by the excited daughter nuclei following
the NC interactions :
\begin{displaymath}
\rm{
\bar{\nu_e} ~ + ~ ^{10}B  ~ \rightarrow ~
\bar{\nu_e} ~ + ~ ^{10}B^*
}
\end{displaymath}
and
\begin{displaymath}
\rm{
\bar{\nu_e} ~ + ~ ^{11}B  ~ \rightarrow ~
\bar{\nu_e} ~ + ~ ^{11}B^* ~ .
}
\end{displaymath}
There are theoretical works~\cite{nuex,nuexaxial}
suggesting that 
these cross sections
are sensitive to the 
axial isoscalar component of NC interactions 
and the strange quark content of the nucleon.
Therefore, $\nu$N NC scattering may provide
a complementary approach to the investigations 
of nucleon structure physics comparing to
the eN scattering systems.
The $\rm{\nu_e}$ NC interaction on $^{11}$B has
been considered as the detection mechanism in the
BOREX solar neutrino proposal~\cite{b11nuex}.

A realistic experiment will consist of about
500~kg of B$_4$C, either in plate or powder
form, inserted into a chamber with CsI crystals
at optimized positions. The experimental
signature will be gamma-lines of the
characteristic energies which show up during
reactor ON period.

If a CsI calorimeter proves
itself to be optimal for studying NC
excitations on nuclei~\cite{nuex,nuexaxial}, 
where the experimental signatures
are the characteristic $\gamma$-lines,
one can insert other passive materials
to measure their cross sections, and
turns the experiment into a longer-term
program.

\subsection{Highlights of Experimental Details}

\begin{figure}
\centerline{
\epsfig{file=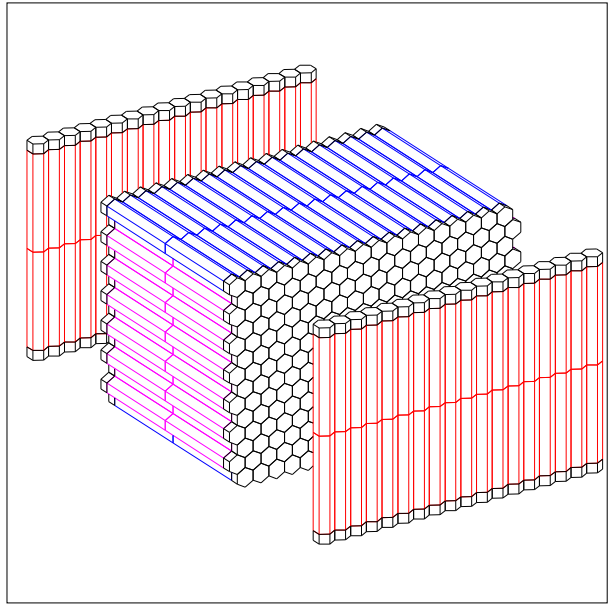,width=8cm}
}
\caption{
Schematic layout of the CsI(Tl) target in
the ``active target'' configuration,
consisting of about 600 crystals,
each of which is hexagonal in shape with 2~cm sides,
20~cm length and 1~kg mass.
}
\label{csitarget}
\end{figure}

Among the various physics items mentioned above,
the first to be pursued will be that of
neutrino-electron scattering, using
the ``active target'' configuration
shown schematically in Figure~\ref{csitarget}.
The detector will consist of about 600~kg of CsI(Tl) 
crystals. Individual crystal
is 1~kg in mass and
hexagonal in shape with  2~cm sides
and 20~cm length. Each channel is read out
by a photo-diode, followed by pre-amplifier,
main amplifier and shaper. The entire pulse
is digitized by a FADC to be read out with
a data acquisition system adopting the VME-bus.

The achieved energy resolution with a 
prototype module is about 16\% FWHM at
660~keV. It is electronic noise-limited and hence
improves linearly with energy. Pulse
shape discrimination between $\gamma$/e and 
$\alpha$ events, as well as those
originated from the
photodiodes, can be achieved to better
than the 99\% level.

\begin{figure}
\centerline{
\epsfig{file=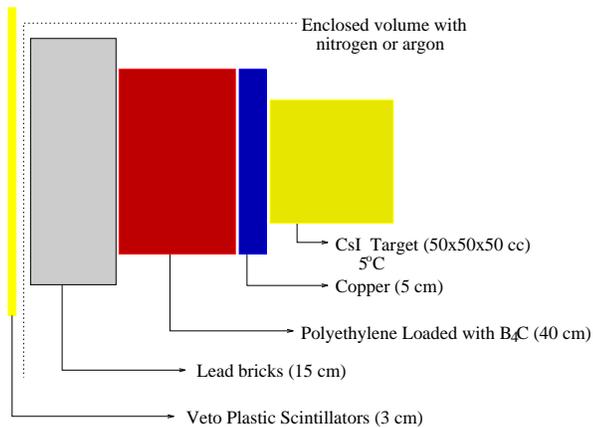,width=8cm}
}
\caption{
Schematic layout of the target and shielding.
The coverage is 4$\pi$ but only one face
is shown. 
}
\label{shielding}
\end{figure}

The CsI target will be shielded by
lead, boron-loaded polyethylene and
copper, as depicted in
Figure~\ref{shielding}. Cosmics will be vetoed by
an outermost layer of plastic scintillators. 
The outer modules of the CsI target can be
used as active veto if necessary. The whole
inner target will be placed in a dry nitrogen
environment to purge the radon gas, and will
be kept at 5$^o$C to reduce electronic noise.

The intrinsic radiopurity level of the CsI(Tl) 
crystal is very crucial to the sensitivities
of this experiment, as well as to 
the future potential applications in low background physics.
By the absence of $\alpha$-peaks above 3~MeV 
in a measurement using a 3~kg crystal in an underground site,
previous work~\cite{csibkg}
have derived that CsI crystals can be grown 
to a purity level such that the concentrations of
$^{238}$U and $^{232}$Th are less
than the $10^{-12}$~g/g level.

A detailed technical report is now under preparation.
We hope to have a first version of the experiment
operational on site by spring 1999.

\section{Outlook}

A Taiwan and mainland China collaboration
has been built up to initiate and pursue
a program in experimental neutrino physics
and astro-particle physics in Taiwan.
A ``pilot'' experiment
to be performed 
close to the reactor core
using CsI(Tl) as detector
is now being prepared.
Various neutrino interactions
at the MeV energy range can be investigated.
The feasibility and conceptual studies 
of the ``next'' project is under way.
A distinct possibility can be a long
baseline reactor neutrino experiment.

This is a pioneering ``foundation''
effort for Taiwan, as well
as the first generation collaborative efforts
in large-scale basic research between scientists
from Taiwan and mainland China.
The importance of  the
outcomes of this experiment and this
experience will 
lie besides, if not beyond, neutrino physics.

\end{document}